\documentclass[reqno]{amsart}
\usepackage{a4wide}

\def\L{{\mathcal L}}
\def\M{{\mathcal M}}

\def\K{{\mathcal K}}

\newtheorem{theorem}{Theorem}
\newtheorem{prop}[theorem]{Proposition}
\newtheorem{lemma}[theorem]{Lemma}

\newcommand{\KK}{{\mathcal K}}
\newcommand{\al}{\alpha}
\newcommand{\ga}{\gamma}

\newcommand{\pal}{\partial}

\begin{document}

\title[Symmetries and Solutions of Getzler's Equation]
{Symmetries and Solutions of Getzler's Equation for
Coxeter and Extended Affine Weyl Frobenius Manifolds}

\author{Ian A.B. Strachan}
\date{$12^{\rm th}$ December 2002}
\address{Department of Mathematics\\ University of Hull\\
Hull HU6 7RX\\ U.K.}

\email{i.a.strachan@hull.ac.uk}


\begin{abstract}
The $G$-function associated to the semisimple Frobenius manifold $
\mathbb{C}^n/W$ (where $W$ is a Coxeter group or an extended
affine Weyl group) is studied. The general form of the
$G$-function is given in terms of a logarithmic singularity over
caustics in the manifold. The main result in this paper is a
universal formula for the $G$-function corresponding to the
Frobenius manifold $\mathbb{C}^n/{\widetilde W}^{(k)}(A_{n-1})\,,$
where ${\widetilde W}^{(k)}(A_{n-1})$ is a certain extended affine
Weyl group (or, equivalently, corresponding to the Hurwitz space
${\hat{M}}_{0;k-1,n-k-1}$), together with the general form of the
$G$-function in terms of data on caustics. Symmetries of the
$G$-function are also studied.
\end{abstract}

\maketitle


\bigskip

\section{Introduction}

The main result in this paper is the following universal formula, independent of $k\,,$
\[
G=-\frac{1}{24} t^n
\]
for the $G$-function corresponding to the Frobenius manifold
$ \mathbb{C}^n/{\widetilde W}^{(k)}(A_{n-1})\,,$ where
${\widetilde W}^{(k)}(A_{n-1})$ is a certain extended affine Weyl group (or,
equivalently, corresponding to the Hurwitz space ${\hat{M}}_{0;k-1,n-k-1}$) \cite{DZ1},
together
with the general form of the $G$-function in terms of data on caustics.
For $n\leq 3$ this result is already known \cite{DZ2}\,, with such solutions being found
by directly solving the governing equations for $G\,.$ For arbitrary $k$ and $n$
a different approach is required.

The $G$-function itself plays a number of important roles within
the mathematics and applications of the theory of Frobenius
manifolds, all connected, especially in the semisimple case, with
the construction of genus one objects from genus zero data. Thus
in TQFT it appears in the genus one contribution to the free
energy of the field theory; in enumerative geometry it governs
genus one Gromov-Witten invariants; in integrable systems it
appears in the first order deformation of bi-Hamiltonian
structures. The $(n=2,k=1)$ case corresponds to the quantum
cohomology of $\mathbb{C}\mathbf{P}^1\,,$ and the precise form of
the $G$-function is required in establishing, to first-order in
the genus expansion, that the underlying integrable system is the
Toda lattice. The results in this paper may to used to construct
similar conjectures for generalized Toda and Benney hierarchies.
It will be assumed in this paper that the reader has an
understanding of Frobenius manifolds. In particular, the concepts
and notation will follow Dubrovin \cite{D}.

In \cite{DZ2} Dubrovin and Zhang (following from conjectures of Givental \cite{Gi1}) proved
that for semisimple Frobenius manifolds the $G$-function is given by the formula
\begin{equation}
G=\log  \frac{ \tau_I}{J^{1/24}}
\label{Gsolution}
\end{equation}
where $\tau_I$ is the isomonodromic $\tau$-function and $J$ is the Jacobian of the
transformation between canonical and flat-coordinates. The governing equations
for $G$ itself were obtained by Getzler \cite{Ge} and are the following overdetermined
set of linear equations \cite{DZ2}:

\begin{equation}
\sum_{1\le \al_1,\al_2,\al_3,\al_4\le n} z_{\al_1} z_{\al_2}
z_{\al_3} z_{\al_4} \Delta_{\al_1\al_2\al_3\al_4}=0
\label{Getz}
\end{equation}
where
\begin{eqnarray*}
\Delta_{\al_1\al_2\al_3\al_4}&=&
3\,c^{\mu}_{\al_1\al_2}\,c^{\nu}_{\al_3\al_4}
\,\frac{\pal^2 G}{\pal t^\mu\pal t^{\nu}}-4\,
c^{\mu}_{\al_1\al_2}\,c^{\nu}_{\al_3\mu}
\,\frac{\pal^2 G}{\pal t^{\al_4}\pal t^{\nu}}
-c^{\mu}_{\al_1\al_2}\,c^{\nu}_{\al_3\al_4\mu}
\,\frac{\pal G}{\pal t^{\nu}}+
\nonumber\\
&& 2\,c^{\mu}_{\al_1\al_2\al_3}\,c^{\nu}_{\al_4\mu} \,\frac{\pal
G}{\pal t^{\nu}}+\frac16
c^{\mu}_{\al_1\al_2\al_3}\,c^{\nu}_{\al_4\mu\nu} +\frac1{24}
c^{\mu}_{\al_1\al_2\al_3\al_4}\,c^{\nu}_{\mu\nu}- \frac14
c^{\mu}_{\al_1\al_2\nu}\,c^{\nu}_{\al_3\al_4\mu}\,.
\end{eqnarray*}
The proof that (\ref{Gsolution}) satisfies these equations involves rewriting them
in terms of canonical coordinates. In particular \cite{DZ2}:
\begin{theorem}
For an arbitrary semisimple Frobenius manifold the system (\ref{Getz}) has a
unique, up to an additive constant, solution $G=G(t^2, \dots, t^n)$
satisfying the quasihomogeneity condition
\begin{equation}
\label{anomaly}
{\mathcal L}_E\,G=\ga
\end{equation}
with a constant $\ga$. This solution is given by the formula (\ref{Gsolution})
where $\tau_I$ is the isomonodromic tau-function and
\[
J=\det \left(\frac{\partial t^\alpha}{\partial u^i} \right)
\]
is the Jacobian of the transform from the canonical coordinates to
the flat ones. The scaling anomaly $\ga$ in (\ref{anomaly}) is given by
the formula
\[
\ga =-\frac{1}{4} \sum_{\alpha=1}^n \mu_\alpha^2 +\frac{n\, d}{48}
\]
where
\[
\mu_\alpha = q_\alpha -\frac{d}{2}, ~\alpha=1, \dots, n.
\]
\end{theorem}
\noindent A simple extension of this result appeared in \cite{DZ3}:

\begin{theorem}
The derivatives of the $G$-function along the
powers
of the Euler vector field are given by the following formulae
\begin{eqnarray}
&&\partial_e G=0,\label{bo7}
\\
&&
\partial_E G= \frac{n\, d}{48} -\frac{1}{4} {\rm tr}\, \mu^2,
\label{bo8}
\\
&&
\partial_{E^k}G =-\frac{1}{4}{\rm tr}\,\left(\mu \, (\mu\, {\mathcal U}^{k-1}
+{\mathcal U}\mu\, {\mathcal U}^{k-2} + \dots + {\mathcal U}^{k-1}\mu)\right)
\nonumber\\
&&\quad
-\frac{1}{24}\left< (\mu\, {\mathcal U}^{k-2}+ {\mathcal U}\mu\, {\mathcal U}^{k-3}+
\dots + {\mathcal U}^{k-2}\mu)\,E -\frac{d}{2}\, {\mathcal U}^{k-2}\,E, H\right>,
\label{bo9}
\\
&&\quad  k\geq 2\nonumber
\end{eqnarray}
where
\[
H=c_\nu^{\nu\alpha}\partial_\alpha.
\]
\end{theorem}

For any given Frobenius manifold one may solve the governing equations given
in the above theorems to find the $G$-function. However, for classes of
Frobenius manifold, such an approach is impractical. The idea behind this
paper is that
it is the singularity structure of these differential equations that
drives the solution: the solution may be derived purely in terms of
data on the singularities. Geometrically these singularities correspond
to caustics in the Frobenius manifold and the singularity data to the $F$-manifold
structure on the caustics.
This idea will be illustrated by finding the $G$-function
for Frobenius manifolds constructed from Coxeter groups and extended-affine Weyl
groups. These caustics also have a number of interesting curvature
properties and applications in integrable systems theory \cite{S}.

\bigskip

The rest of the paper is laid out as follows.  In section 2 the
relationship between the multiplication on caustics and the
singularities of the $G$-function is studied. In section 3 these
results will be used to study the $G$-function for the Frobenius
manifold $\mathbb{C}^n/{W}\,,$ where $W$ is a Coxeter group. In
section 4 the $G$-function will be studied for the Frobenius
manifold $\mathbb{C}^n/{\widetilde W}\,,$ where $\widetilde W$ is
an extended affine Weyl group, concentrating in particular on the
special case ${\widetilde W}={\widetilde W}^{(k)}(A_{n-1})\,.$
Frobenius manifolds have certain natural symmetries \cite{D}, and
such symmetries induce symmetries of the corresponding
$G$-functions. Such symmetries are studied in section 5.

\section{Multiplication on caustics and the $G$-function}

By definition, a massive Frobenius manifold $\M$ has a semisimple multiplication
on the tangent space at {\sl generic} points of $\M\,.$ The set of points where the multiplication
is not semisimple is known as the caustic, and will be denoted $\K\,.$ This is an
analytic hypersurface in $\M\,,$ which may consist of a number of
components (possibly highly singular),
\[
\K = \bigcup_{i=1}^{\#\K_i} \K_i\,.
\]
The set of smooth points in $\K$ will be denoted $\K_{reg}\,.$

The simplest case case is where the multiplication on the caustic $\K_i$ is
of the type $A_1^{n-2} \,I_2(N_i)\,,$ i.e. the multiplication decomposes into
$n-2$ one-dimensional algebras and a single two-dimensional algebra
based on the Coxeter group $I_2(N)\,.$ The following theorem
studies the behaviour of the $G$-function near such caustics.

\begin{theorem}\cite{He}
Let $(M,\circ,e,E,g)$ be a simply connected massive Frobenius manifold.
Suppose that at generic points of the caustic $\KK_i$ the germ of the
underlying F-manifold is of type $I_2(N_i)A_1^{n-2}$ for one
fixed number $N_i\geq 3$.

a) The form $d\, \log \tau_I$ has a logarithmic pole along $\KK_i$ with
residue $-\frac{(N_i-2)^2}{16 N_i}$ along $\K_i\cap\KK_{reg}$.

b) The form $-\frac{1}{24} d \log J$ has a logarithmic pole along $\KK_i$
with residue $\frac{N_i-2}{48}$ along $\K_i\cap\KK_{reg}$.

c) The G-function extends holomorphically over $\KK_i$ iff $N_i=3$.
\end{theorem}

\noindent The explicit form of the multiplication used in the
proof of this theorem is, in terms of the coordinate fields
$\delta_i=\frac{\partial~}{\partial t^i}\,,$ with respect to some
not necessarily flat coordinates $t^i\,,$
\begin{equation}
\begin{array}{rcl}
\delta_1 \circ \delta_2 & = & \delta_2 \,, \\
\delta_2 \circ \delta_2 & = & (t^2)^{N-2} \, \delta_1 \,,\\
\delta_i \circ \delta_j & = & \delta_{ij} \, \delta_j\,\qquad\qquad {\rm otherwise}\,,
\end{array}
\label{multiplication}
\end{equation}
with caustic $\KK=\{{\bf t}\,|\,t^2=0\}\,.$ The canonical
coordinates are, on a simply connected subset of $M-\KK\,,$
\begin{eqnarray*}
u_1 & = & t^1 + \frac{2}{N} (t^2)^\frac{N}{2}\,,\\
u_2 & = & t^1 - \frac{2}{N} (t^2)^\frac{N}{2}\,,\\
u_i & = & t^i \,, \qquad\qquad i\geq 3\,,
\end{eqnarray*}
the idempotent vector fields are
\begin{eqnarray*}
e_1 & = & \frac{1}{2} \delta_1 + \frac{1}{2} {(t^2)}^{-\frac{N-2}{2}} \, \delta_2\,,\\
e_2 & = & \frac{1}{2} \delta_1 - \frac{1}{2} {(t^2)}^{-\frac{N-2}{2}} \, \delta_2\,,\\
e_i & = & \delta_i \,, \qquad\qquad i\geq 3
\end{eqnarray*}
and the Euler vector field is
\[
E=t^1 \delta_1 + \frac{2}{N} t^2 \delta_2 + \sum_{i=3}^n t^i
\delta_i\,.
\]
With these one may directly calculate the forms $d\log\tau_I$ and
$d\log J$ near the caustic to obtain the above result. However,
not all caustics are of this simple kind, as the following example
shows.

\medskip

\noindent {\bf Example~} Consider the Frobenius manifold $Q_r(\mathbb{C}\mathbf{P}^1)\,,$ related to the quantum cohomology of
$\mathbb{C}\mathbf{P}^1$, given by the prepotential
and Euler field
\begin{eqnarray*}
F&=&\frac{1}{2}(t^1)^2 t^2 + e^{r t^2}\,,\quad (r>0)\,,\\
E&=&t^1 \partial_1 + \frac{2}{r} \partial_2\,.
\end{eqnarray*}
The canonical coordinates are easily calculated:
\begin{eqnarray*}
u_1 & = & t^1 + 2 r^{\frac{1}{2}} e^{\frac{r}{2} t^2}\,,\\
u_2 & = & t^1 - 2 r^{\frac{1}{2}} e^{\frac{r}{2} t^2}\,.
\end{eqnarray*}
Note that $u_1-u_2$ does not vanish at finite points, but will vanish in the
limit $t^2 \rightarrow -\infty.$ Thus one has a \lq caustic at infinity\rq~or
limiting caustic
\[
\K_\infty=\{ (t^1,t^2) \,|\, t^2\rightarrow -\infty\}\,.
\]
This idea of a \lq caustic at infinity\rq~may be made more precise by introducing
new coordinates
\begin{eqnarray*}
{\tilde t}^1 & = & t^1\,,\\
{\tilde t}^2 & = & e^{t^2}\,.
\end{eqnarray*}
In these coordinates the caustic becomes
$\K_{log}=\{ ({\tilde t}^1,{\tilde t}^2) \, |\, {\tilde t}^2=0\}\,$
and the multiplication takes the form (with $\bf t$ replaced by ${\bf\tilde t}$)
given in the first two equations
of (\ref{multiplication}) with $r=N$. Such a caustic will be referred to as a
logarithmic caustic.
These new variables are no longer flat - the metric is logarithmic along $\K_{log}\,,$
\begin{eqnarray*}
g & = & 2 \, dt^1 \, dt^2 \,, \\
& = & 2 \, d{\tilde t}^1 \, \frac{d{\tilde t}^2}{{\tilde t}^2}\,,
\end{eqnarray*}
but the $F$-manifold structure (as the next Lemma will show) is easier to
understand in these coordinates.

\bigskip

It is illuminating to calculate the $G$-function purely in terms of canonical
coordinates, as this will mirror the calculations in the next lemma.
The $G$-function is made up of two parts (see \cite{DZ2,DZ3}):
\begin{eqnarray*}
d\log\tau_I & = & \frac{1}{8} (u_1-u_2) \frac{\eta_{12}^2}{\eta_1\eta_2} \, d(u_1-u_2)\,,\\
& = & -\frac{r}{16}\, \frac{d{\tilde t}^2}{{\tilde t}^2}
\end{eqnarray*}
(using the fact that the Egoroff potential $\eta=t^2$) and
\[
d\log J = -\frac{r}{2} \frac{d{\tilde t}^2}{{\tilde t}^2}\,.
\]
Hence from (\ref{Gsolution}) $dG=-\frac{r}{24}\, \frac{d{\tilde t}^2}{{\tilde t}^2}$
and since $G$ is independent of $t^1\,,$
$G=-\frac{r}{24} \log{\tilde t}^2\,.$ Note that $dG$ has a logarithmic pole
along $\K_{log}\,$
(In fact, for any two-dimensional semisimple Frobenius manifold the scaling
anomaly and the equations
\begin{eqnarray*}
\L_e G & = & 0 \,, \\
\L_E G & = & \gamma
\end{eqnarray*}
determine $G$ up to a constant, $G=\gamma\log(u_1-u_2)\,.$ From
this one sees the close relationship between properties of the
$G$-function and caustics).

\bigskip

The next result studies the behaviour of the $G$-function near such a limiting caustic.
The proof is almost identical to the proof \cite{He} of theorem 3.

\begin{lemma}
Let $M\subset \mathbb{C}^n$ be a manifold with coordinates ${\tilde t}^1,...,{\tilde t}^n$
and $\K_{log} := \{{\bf \tilde t}\,|\,{\tilde t}^2=0\}$ such that $\M-\K_{log}$
is a Frobenius manifold $(\M-\K_{log},\circ,e,E,g)$ with the following properties:
for some $N_{log}\geq 1$ the multiplication $\circ$ is given by
\[
\begin{array}{rcl}
\delta_1 \circ \delta_2 & = & \delta_2 \,, \\
\delta_2 \circ \delta_2 & = & ({\tilde t}^2)^{N_{log}-2} \, \delta_1 \,,\\
\delta_i \circ \delta_j & = & \delta_{ij} \, \delta_j\,\qquad\qquad {\rm otherwise}\,,
\end{array}
\]
where $\delta_i=\frac{\partial~}{\partial {\tilde t}^i}$ (then for $N_{log}\geq 2$
it extends holomorphically to $\K_{log}$ and for $N_{log}\geq 3$ $\K_{log}$
is the caustic); the metric is logarithmic along $\K_{log}$ , i.e. the matrix of
components of the metric $g$ for a base of logarithmic vector
fields, with respect to $\K_{log}\,,$ is holomorphic and
nondegenerate on the caustic $\K_{log}\,$. Then near the caustic
$\K_{log}\,,$
\begin{eqnarray*}
d \log J & = & -\frac{N_{log}}{16} \, \frac{d{\tilde t}^2}{{\tilde t}^2}+
{\rm holomorphic~one~form~in~}{\bf \tilde t}\,,\\
d \log \tau_I & = & -\frac{N_{log}}{2}  \, \frac{d{\tilde t}^2}{{\tilde t}^2}+
{\rm holomorphic~one~form~in~}{\bf \tilde t}\,.
\end{eqnarray*}

\noindent and hence
\[
dG = -\frac{N_{log}}{24}\,  \frac{d{\tilde t}^2}{{\tilde t}^2}+
{\rm holomorphic~one~form~in~}{\bf \tilde t}\,.
\]
\end{lemma}

\bigskip

\noindent It will be shown in section 4 that for Frobenius manifolds constructed
from extended affine Weyl groups the assumption made in the above lemma holds. The
lemma shows that near the caustic $\K_{log}$ the form $dG$ is,
in the original flat coordinates, finite. This
fact may then be used to exclude the possibility of terms like $e^{-t^2}$ appearing in
the $G$-function.

\bigskip

\noindent{\bf Proof~} Note for $N_{log}\geq 2$ the $F$-manifold
structure is of type $I_2(N_{log})A_1^{n-2}$ (and $I_2(2)=A_1^2$).
For $N_{log}=1$ the multiplication  has a simple pole along
$\K_{log}\,.$ The proof is entirely analogous to the proof of
Theorem 3. The socle field is
\[
H= 2 {\tilde t}^2 \delta_2 + \delta_3 + \ldots + \delta_n\,,
\]
and now ${\tilde t}^2 \delta_2 (\eta)(0)\neq 0\,.$
With this data one can repeat the proof of Theorem 3, as given in \cite{He}, to derive the
result.

\section{The $G$-function for Coxeter groups}

The construction of a Frobenius manifold structure on the orbit space $ \mathbb{C}^n/W$
(where $W$ is a Coxeter group) is given in \cite{D}. The only parts of that construction that
will be required in this section are the following:

\begin{itemize}
\item{}
in flat-coordinates, the prepotential, and hence
the structure functions of the Frobenius algebra, are polynomial functions;
\item{}
the Euler vector field takes the form
\[
E=\sum_{r=1}^n\frac{d_r}{h} t^r \frac{\partial~}{\partial
t^r}\,\qquad\qquad d_r>0
\]
where the $d_r$ are the exponents of the Coxeter group and $h$ is the Coxeter number of $W\,.$
\end{itemize}

\noindent The components of the caustic $\KK$ for such orbit spaces are given in terms of
quasihomogeneous polynomials $\kappa_i$ such that $\kappa_i^{-1}(0)=\K_i\,.$ The $F$-manifold
structure on these caustics is known, the multiplication is of type $I_2(N_i) A_1^{n-2}\,,$
and this enables Theorem 3 to be used. The data $N_i$ is given in Table 1.
It can be extracted with some work from \cite{H2} Theorem 5.22, which builds on
\cite{Gi2}.

\begin{prop} The $G$-function on $ \mathbb{C}^n/W$ takes the form
\begin{equation}
G=-\frac{1}{24} \frac{(N_1-2)(N_1-3)}{N_1} \log \kappa_1
\label{GCoxeter}
\end{equation}
and the constant $N_1\,,$ which depends on the Coxeter group
$W\,,$ is given in table 1.
\end{prop}

\bigskip

\begin{table}
\begin{center}
\begin{tabular}{c|c|c}
Coxeter Group $W$ & Number of caustics & Values of $N_i$ \\ \hline
$A_n\,,D_n\,,E_{6,7,8}$ & $1$ & $N_1=3$ \\
& & \\
$B_n$ & $2$ & $N_1=4\,,N_2=3$\\
& & \\
$F_4$ & $3$ & $N_1=4\,,N_2=N_3=3$\\
& & \\
$H_3$ & $2$ & $N_1=5\,,N_2=3$\\
& & \\
$H_4$ & $2$ & $N_1=5\,,N_2=3$\\
& & \\
$I_2(h)$ & $1$ & $N_1=k$\\
\end{tabular}
\end{center}
\vskip 5mm
\caption{Data on the caustics of the Frobenius manifold $\mathbb{C}^n/W$}
\end{table}

\noindent{\bf Proof~} From theorem 2 it follows that the only singularities of $dG$
are on caustics, and it is known that the multiplication on the caustics
of a Frobenius manifold obtained from a Coxeter group is of the form where Theorem 3
may be applied (see \cite{Gi2} and \cite{H2} Theorem 5.22).

It follows from the polynomial nature of the structure functions
and Theorem 2 that all first derivatives $\partial G/\partial t^\alpha$ are rational
functions. Hence, on integrating, $G$ takes the schematic form
\[
G(t)={\rm rational~function~} + {\rm logarithmic~singularities}
\]
By Theorem 3, the only singularities that $G$ has are logarithmic singularities on $\KK_i\,.$
Thus the rational functions must be polynomial. However, the only polynomial
function compatible with the symmetry (\ref{anomaly}) is a constant (this uses the fact that
the exponents of the Coxeter group are all positive). Since $G$ is only defined up
to a constant anyway one has:
\begin{equation}
G(t) = -\frac{1}{24}\sum_{i=1}^{\#\KK_i} \frac{(N_i-2)(N_i-3)}{N_i}\log\kappa_i
\label{GpreGivental}
\end{equation}
and hence
\begin{equation}
\gamma = -\frac{1}{24}\sum_{i=1}^{\#\KK_i} \frac{(N_i-2)(N_i-3)}{N_i}E\left(\log\kappa_i\right)\,.
\label{gammacox}
\end{equation}
Using the data in table 1,
and in particular that in all cases
there is at most one caustic, denoted $\K_1\,,$ with $N_i>3\,,$ the result follows.

\bigskip

The scaling constant $\gamma$ may be calculated purely from the data on the caustics:
\begin{equation}
\gamma=-\frac{1}{24} \frac{(N_1-2)(N_1-3)}{N_1}
E(\log(\kappa_1))\,.
\end{equation}
so in
particular for $W=B_n\,,$
\[
E( \kappa_1^{B_n}) = \left(\frac{n-1}{n}\right) \, \kappa_1^{B_n}\,.
\]
The solutions are summarized in Table 2. This uses the explicit form of the prepotential
for various 4-dimensional Frobenius manifolds given in \cite{D2}. The constants $\gamma$ are easily
found using the exponents of the Coxeter group. Note that if $\gamma=0$ then one may
deduce from (\ref{gammacox}) that $N_i=3$ for all $i$ and hence $G=0\,,$ even without
knowing the number of components of the caustic.

\bigskip

\begin{table}
\begin{center}
\begin{tabular}{c|c|c}
Coxeter Group $W$ & $\gamma_{_W}$ & $G_W$ \\ \hline
$A_n$ & $0$ & $0$ \\
& & \\
$B_n$ & $\frac{(1-n)}{48 n}$ & $ -\frac{1}{48} \log \kappa_1^{B_n}$\\
& & \\
$D_n$ & $0$ & $0$ \\
& & \\
$E_{6,7,8}$ & $0$ & $0$ \\
& & \\
$F_4$ & $-\frac{1}{48}$ & $-\frac{1}{48} \log[ 6 t_3^2 - 2 t_2 t_4^2 + t_4^6]$ \\
& & \\
$H_3$ & $-\frac{3}{100}$ & $-\frac{1}{20} \log[t_2 -  t_3^3]$\\
& & \\
$H_4$ & $-\frac{1}{25}$ & $-\frac{1}{20} \log[2025 t_3^2-8100 t_2 t_4^2 + 630 t_3 t_4^6 - 16 t_4^{12}]$\\
& & \\
$I_2(h)$ & $-\frac{1}{12}\frac{(h-2)(h-3)}{h^2}$ & $-\frac{1}{24}\frac{(2-h)(3-h)}{h} \log[ t_2]$\\
\end{tabular}
\end{center}
\vskip 5mm
\caption{The $G$-function on the space $\mathbb{C}^n/W$}
\end{table}

\vskip 5mm

\section{The $G$-function for extended affine Weyl groups}

The construction of a Frobenius manifold structure on the orbit space $ \mathbb{C}^n/{\widetilde W}$
(where ${\widetilde W}$ is an extended affine Weyl group) is given in \cite{DZ1}. The
following parts of their construction
will be required in this section:

\begin{itemize}
\item{}
in flat coordinates, the prepotential, and hence
the structure functions of the Frobenius manifold, are polynomial functions in
$\{t^1\,,t^2\,,\ldots\,,t^{n-1}\,,e^{t^n}\}\,;$
\item{}
the Euler vector field takes the form
\[
E=\sum_{r=1}^{n-1}\frac{d_r}{d_k} t^r \frac{\partial~}{\partial t^r}+
\frac{1}{d_k}  \frac{\partial~}{\partial t^n}\,\qquad\qquad d_r>0
\]
where the $d_r$ are various numbers related to the extended affine Weyl groups, which may be
found in Table 2 of \cite{DZ1}\,.
\end{itemize}
In addition one requires the following properties of caustics and limiting caustics
for these manifolds:

\begin{lemma} The extended affine Weyl group Frobenius manifolds
are coverings of Frobenius manifolds
\[\M-\K_{log}=\mathbb{C}^n-\{ {\bf {\tilde t}}\,|\, {\tilde t}^n =0\}\,,
\]
with covering map
${\bf t}\mapsto {\bf {\tilde t}} = \{t^1\,,\ldots\,,t^{n-1}\,,e^{t^n}\}\,.$
At generic points of $\K_{log}$ the assumptions of Lemma 4 are satisfied.
The caustics of the extended affine Weyl group Frobenius manifolds
are given by the vanishing of certain quasihomogeneous polynomials in
the variables $\{t^1\,,t^2\,,\ldots\,,t^{n-1}\,,e^{t^n}\}\,.$
\end{lemma}

\noindent{\bf Proof } The proof is simpler in flat coordinates $\{ {\bf t} \}$ rather than
the $\{ \bf {\tilde t} \}$ coordinates. The canonical coordinates are the roots of the polynomial
$poly(\lambda)=0$ where
\[
poly(\lambda) = \det[g^{ij}(t) - \lambda \eta^{ij}(t)]\,.
\]
Using the information contained within the details of Lemma 2.6 in \cite{DZ1} (in particular
the equation preceding  (2.30)) it is easy to show that
\[
\lim_{t^n\rightarrow -\infty} g^{kr} = 0
\]
for $r\neq n\,$ (recall that for these manifolds the identity
element is given by $e=\frac{\partial~}{\partial t^k}$ rather than
$e=\frac{\partial~}{\partial t^1}$). This shows, on expanding the
determinant along the $k^{\rm th}$ row or column, that in the
limit $t^n\rightarrow - \infty\,,$ the polynomial $poly(\lambda)$
has a repeated root. Hence $\K_\infty$ is a limiting caustic, or
equivalently, $\K_{log}$ is a logarithmic caustic. Consideration
of the resultant of $poly(\lambda)$ also shows that the standard
caustics are all given in terms of quasihomogeneous polynomials
$\kappa_i\,,$ via $\K_i = \kappa_i^{-1}(0)\,,$ where the
$\kappa_i$ are polynomial in the variables $\{ {\bf \tilde
t}\}\,.$ In addition, the Egoroff potential for these manifolds is
given by
\[
\eta=t_k = \eta_{kr} t^r = t^n\,.
\]
Thus $\left.({\tilde t}^n \delta_n \eta)\right|_{\K_{log}}$ is
constant. Thus the assumptions in lemma 4 hold for these
manifolds.

\begin{prop} Under the assumption that the multiplication on the caustics
$\K_i$ of the Frobenius manifold $\mathbb{C}^n/{\widetilde W}$ is
of the form $A_1^{n-2} I_2(N_i)$ the associated $G$-function is
\begin{equation}
G=-\frac{N_{log}}{24} \, t^n-\frac{1}{24} \sum_{i=1}^{\#\K_i}
\frac{(N_i-2)(N_i-3)}{N_i} \log \kappa_i \label{GEAW}
\end{equation}
where
\[
\gamma = -\frac{N_{log}}{24\, d_k} - \frac{1}{24}
\sum_{i=1}^{\#\K_i} \frac{(N_i-2)(N_i-3)}{N_i}
E\left(\log\kappa_i\right)\,.
\]
\end{prop}

\bigskip

\noindent{\bf Proof~} The idea behind this proof is similar to the
one used in proposition 5. One is integrating rational functions
so the $G$-function has only rational and logarithmic terms. Under
the above assumption one can exclude pole singularities and by
lemma 4 one can exclude behaviour like $e^{-t^n}\,.$ A scaling
argument then forces a polynomial to be a constant.

In more detail, the argument runs as follows. In flat coordinates
equations (\ref{bo7}),(\ref{bo8}) and (\ref{bo9}) (for $k=2\,,\ldots\,,n$) may be inverted to
find all the first derivatives of the $G$-function with respect to the flat
coordinates. These must be rational functions with the same denominator, and so may be written
\begin{equation}
\frac{\partial G}{\partial t^i} = \frac{p_i(t)}{\Delta(t)}\,, \quad\quad i=1\,,\ldots\,, n\,,
\label{Gderivatives}
\end{equation}
where $p_i$ and $\Delta$ are quasihomogeneous polynomials in
$\{t^1\,,\ldots\,,t^{n-1}\,,e^{t^n}\}\,.$
It is useful to introduce a slightly different set of variables,
$\{{\tilde t}^\alpha\}$, defined by ${\tilde t}^\alpha=t^\alpha\,, \alpha=1\,,\ldots\,,n-1\,,$
and ${\tilde t}^n=e^{t^n}\,.$ In these new variables (\ref{Gderivatives})
become
\[
\frac{\partial G}{\partial {\tilde t}^i} =
\frac{ {\tilde p}({\tilde t})}{{\tilde t}^n \, \Delta({\tilde t})}
\]
where the extra term in the denominator comes from the chain rule. On integrating the singularities
will come from the zeroes of the denominator. Thus on integrating
\[
G({\tilde t})={\rm ~rational~function~} + {\rm logarithmic~singularities}\,.
\]
Under the assumption and theorem 3, together with lemma 4 to exclude terms involving
$e^{-t^n}\,,$ the rational function must be polynomial and the scaling argument
implies that this is then a constant. Thus,
\[
G({\tilde t})=-\frac{N_{log}}{24} \log {\tilde t}^n - \frac{1}{24}
\sum_{i=1}^{\#\K_i} \frac{(N_i-2)(N_i-3)}{N_i} \log
\kappa_i({\tilde t})
\]
which becomes (\ref{GEAW}) on converting back to the flat variables. Application of the
Euler vector field then gives the final part of the lemma.

\bigskip

Note, this is just the conjectural form of the $G$-function; in order to prove it one
requires information on the $F$-manifold structure on the caustics of these manifolds.
In the special case of the group ${\widetilde W}={\widetilde W}^{(k)}(A_{n-1})$ one
may easily derive, using ideas from singularity theory, the required $F$-manifold structure and hence the explicit form
of the $G$-function.

\begin{theorem} For the Frobenius manifold
$\mathbb{C}^n/{\widetilde W}^{(k)}(A_{n-1})$ the $G$-function takes the
universal form
\begin{equation}
G=-\frac{1}{24} t^n\,,
\label{mainresult}
\end{equation}
independent of $k\,.$

\end{theorem}

\noindent{\bf Proof~} From the above lemma it suffices to show that the manifold $\mathbb{C}^n/{\widetilde W}^{(k)}(A_{n-1})$
has only one caustic $\K_1$ (together with the limiting caustic $\K_\infty$) on which the
multiplication is of the form $A_1^{n-2} I_2(3)\,.$ The scaling anomaly may be
derived from the known data (see table 3).

The proof that $N_1=3$ comes from a standard argument in singularity theory. The manifold
$\M$ is isomorphic to the base space of the unfolding
\[
F_{\bf a}(x) = x^k + a_1 x^{k-1} + \ldots + a_k + \ldots + a_{k+m} x^{-m}
\]
with parameters $(a_1\,,\ldots\,, a_{k+m}) \in \mathbb{C}^{k+m-1} \times \mathbb{C}^\ast$
and variable $x \in \mathbb{C}^\ast$ (in \cite{DZ1} $x=e^{i\phi}$). The total space of
the unfolding has a critical space $C\subset \mathbb{C}^\ast \times \M$ (of dimension
$n=k+m$) and the projection $C\rightarrow \M$ is a branched covering with the caustic $\K$
being the image of the critical points of the projection. In this case the critical space
is smooth and the set of critical points of the projection $C\rightarrow \M$ is also
smooth: from this it follows that the caustic $\K$ has only one component and that
$N_1=3\,.$ The same proof may be used to derive the data in the first line of table 1 for
the Coxeter group $A_n\,.$

\bigskip

\noindent A plausible conjecture is that the assumption in
proposition 7 is true and also that the data on the caustics of
the extended affine Weyl groups is the same as the corresponding
Coxeter groups. For $B_3\,,C_3\,,G_2$ extended affine Weyl groups
this may be verified by direct calculation using the explicit
formulae for their prepotentials given in \cite{DZ1}. If the
conjecture is true then $G=-1/24 \,t^n$ for the $D_l$ and
$E_{6,7,8}$ extended affine Weyl groups as well.

\begin{table}
\begin{center}
\begin{tabular}{c|c}
$\widetilde W$ & $\gamma_{_{\widetilde W}}$ \\ \hline
$A_l^{(k)}\,,D_l\,,E_{6,7,8}$ & $-\frac{1}{24} \frac{1}{d_k}$\\
 & \\
$B_l$ & $-\frac{1}{48} \frac{(l+1)}{d_k}$\\& \\
$C_l$ & $-\frac{1}{24} \frac{(l+1)}{d_k}$\\& \\
$F_4$ & $-\frac{5}{144} $\\ & \\
$G_2$ & $ - \frac{1}{16}$ \\
 & \\
\end{tabular}
\end{center}
\vskip 5mm
\caption{The scaling anomalies of the Frobenius manifolds $\mathbb{C}^{l+1}/{\widetilde W}$}
\end{table}

\vskip 5mm

\section{Symmetries of the $G$-function}

Symmetries of the WDVV equation are transformations
\begin{eqnarray*}
t^\alpha & \mapsto & {\hat t}^\alpha\,, \\
\eta_{\alpha\beta} & \mapsto & {\hat\eta}_{\alpha\beta}\,, \\
F & \mapsto & {\hat F}
\end{eqnarray*}
which preserve the equations. In appendix B in \cite{D} two types of symmetries
were described, a Legendre-type transformation $S_\kappa$ and an inversion $I\,.$
Such transformations are defined in terms of flat-coordinates, but in terms of canonical
coordinates and isomonodromic data they are very simple, the details being the content
of Lemma 3.13 and Proposition 3.14 in \cite{D}\,. Such transformations induce
transformations in the isomonodromic $\tau_I$ functions (basically Schlesinger transformations)
and the $G$-function:

\begin{lemma} Under a symmetry $S_\kappa\,:$
\begin{eqnarray*}
{\hat \tau}_I & = & \tau_I\,,\\
{\hat G} & = & G - \frac{1}{24} \log \det\left(
\frac{\partial ({\hat t}^1,\ldots,{\hat t}^n)}{\partial ({t}^1,\ldots,{t}^n)}\right)\,,\\
{\hat\gamma} & = & \gamma - \frac{n q_\kappa}{24}\,.
\end{eqnarray*}
\bigskip
Under an inversion $I\,:$

\begin{eqnarray*}
{\hat \tau}_I & = & \frac{\tau_I}{\sqrt{t^n}}\,,\\
{\hat G} & = & G +\left(\frac{n}{24}-\frac{1}{2}\right) \log t^n\,,\\
{\hat\gamma} & = & \gamma + \left( \frac{n}{24}-\frac{1}{2}\right) (1-d)\,.
\end{eqnarray*}
(This second result assumes that
the Euler vector field takes the form
\[
E=\sum_{i=1}^n d_i t^i \frac{\partial~}{\partial t^i}
\]
and that $d\neq 1$. The result may be easily extended
to cover more general cases).
\end{lemma}

\bigskip

\noindent No proof will be given - it follows immediately from
Lemma 3.13 and Proposition 3.14 in \cite{D} together with the transformation properties
of Jacobians. These formulae may be used to define the $G$-function
on twisted Frobenius manifolds. Note also the invariant properties of
$G$ under inversions if $n=12\,.$

\bigskip

\noindent{\bf Example} Consider the prepotential corresponding to the
extended affine Weyl group $A^{(1)}_2\,:$

\[
F=\frac{1}{2} t_1^2 t_3 + \frac{1}{2} t_1 t_2^2 - \frac{1}{24} t_2^4 + t_2 e^{t_3}\,.
\]
The corresponding $G$-function is given by (\ref{mainresult}):
\[
G=-\frac{1}{24}t^3\,.
\]
Under the symmetry $S_2$ and $S_3$ one obtains, respectively, the prepotentials
\begin{eqnarray*}
{ \hat F} & = & \frac{1}{6} {\hat t}_2^3 +  {\hat t}_1  {\hat t}_2 {\hat t}_3 +
\frac{1}{6} {\hat t}_1 {\hat t}_3^3 + \frac{1}{2} {\hat t}_1^2 \left( \log {\hat t}_1 -
\frac{3}{4}\right)\,,\\
{\hat F} & = & \frac{1}{2} {\hat t}_1 {\hat t}_3^2 +
\frac{1}{2} {\hat t}_2^2{\hat t}_3 + \frac{1}{2} {\hat t}_1^2 \log {\hat t}_2\,.
\end{eqnarray*}
Application of the above lemma then gives the following $G$-functions:
\begin{eqnarray*}
{\hat G} & = & - \frac{1}{12} \log {\hat t}_1 \,, \\
{\hat G} & = & - \frac{1}{8} \log {\hat t}_2\,.
\end{eqnarray*}
The first of these solutions was obtained in \cite{CT} by directly solving
Getzler's equations (\ref{Getz}) from the corresponding prepotential.
The results of this paper may be used to
construct the genus one corrections to the bi-Hamiltonian hierarchy for
generalized, multi-component, Toda and Benney hierarchies.

\section*{Appendix}

In this appendix some of the results of this paper will be extended to the Hurwitz
space $H_{1,0;l}\,,$ which coincides with the orbit space $ \mathbb{C}^{l+1}/J(A_l)$
where $J(A_l)$ is a Jacobi group - a particular extension of complex crystallographic
groups \cite{D,B}. By direct calculation the $G$-functions for the first two
members of this series are
\begin{eqnarray*}
G_{J(A_1)} & = & -\frac{1}{\pi^4} \log \eta{(t_0)} - \frac{1}{8} \log t_3\,, \\
G_{J(A_2)} & = & -\frac{1}{\pi^4} \log \eta{(t_0)} - \frac{1}{6} \log t_4\,,
\end{eqnarray*}
where $\eta$ is the Dedekind function (for the corresponding prepotentials, see \cite{B}).
The scaling anomaly is easily calculated in general,
\[
\gamma_{J(A_l)} = -\frac{1}{24} \frac{l+2}{l+1}\,,
\]
which suggests the conjectural form
\[
G_{J(A_l)}  =  -\frac{1}{\pi^4} \log \eta{(t_0)} - \frac{l+2}{24} \log t_{l+2}\,.
\]
Related results have appeared recently in \cite{KK}. It would be
interesting to see how these approaches compare. These
calculations suggest that the $G$-function for these spaces should
have a simple form, though a deeper understanding of the
properties of caustics in these spaces will be required in order
to apply the methods of this paper.

\medskip

\noindent{\bf Acknowledgment:} I would like to thank Claus Hertling for explaining
certain details of \cite{He} to me and for his detailed criticism of an earlier version
of this paper. Financial support was provided by the EPSRC, grant GR/R05093.

\end{document}